\documentclass[prl,aps,twocolumn,reprint,floatfix,nofootinbib,superscriptaddress]{revtex4-1}
\usepackage[pdftex]{hyperref}
\usepackage{url}

\usepackage{amssymb,slashed,latexsym,ifpdf,amsmath}

\def\K{K{\"a}hler}
\newcommand{\BPhi}{\mathbf{\Phi}}
\newcommand{\BPhibar}{\mathbf{\bar \Phi}}
\newcommand{\BS}{\mathbf{S}}
\newcommand{\BSbar}{\mathbf{\bar S}}

\newcommand{\BT}{\mathbf{T}}
\newcommand{\BTbar}{\mathbf{\bar T}}

\newcommand{\BZ}{\mathbf{Z}}
\newcommand{\BZbar}{\mathbf{\bar Z}}

\newcommand{\BA}{\mathbf{A}}
\newcommand{\BB}{\mathbf{B}}

\def\be{\begin{equation}}
\def\ee{\end{equation}}
\def\ba{\begin{array}}
\def\ea{\end{array}}
\def\bea{\begin{eqnarray}}
\def\eea{\end{eqnarray}}

\newcommand{\rf}[1]{(\ref{#1})}
\newcommand{\vp}{\varphi}

\begin{document}

\title{\Large Inflatino-less Cosmology}
\author{John Joseph M. Carrasco}
\affiliation{Institut de Physique Th\'eorique,
CEA/DSM/IPhT, 
CEA-Saclay, 
 91191 Gif-sur-Yvette, France}
 \affiliation{School of Natural Sciences, Institute for Advanced Study, Princeton, NJ, USA}
\author{Renata Kallosh}
\author{Andrei Linde}
\affiliation{SITP and Department of Physics, Stanford University, \\
Stanford, CA
94305 USA}

\begin{abstract}
We construct  inflationary models in the context of  supergravity  with orthogonal nilpotent superfields \cite{FKT}. When local supersymmetry is gauge-fixed in the unitary gauge, these models describe theories with only a single real scalar (the inflaton), a graviton and a gravitino. Critically, there is no inflatino, no sgoldstino, and no sinflaton  in these models. This dramatically simplifies  cosmological models which can simultaneously describe inflation, dark energy and SUSY breaking.
\end{abstract}

\maketitle

\noindent{\bf{Introduction.}}
\label{intro}
In this letter we show that the recently constructed models of supergravity with orthogonal nilpotent superfields  \cite{FKT}  significantly simplify  construction of cosmological models which simultaneously describe inflation, dark energy and SUSY breaking. Now one can achieve this goal using the absolutely minimal number of ingredients:  graviton,  gravitino, and a single real scalar, the inflaton. Our methods apply to a broad class of inflationary theories, but they are especially suitable for describing  $\alpha$-attractors \cite{Kallosh:2013hoa}, which provide a very good fit to the latest cosmological data \cite{Ade:2015lrj}.

\vskip 3pt 
\noindent{\bf{Orthogonal nilpotent superfields.}}
Global supersymmetry models with orthogonal nilpotent superfields  were studied in \cite{Komargodski:2009rz,Kahn:2015mla}.
 Their generalization to local supergravity interacting with orthogonal nilpotent multiplets was presented
in \cite{FKT}.  The models depend on two constrained chiral superfields\footnote{We use bold face letters  for superfields which in global case are functions of $(x, \theta,\bar \theta) $. For example, the chiral superfield $\BS$ means that $\BS = S(x) + \sqrt 2 \theta \chi^s + \theta^2 F^s$, and $\BPhi = \Phi(x) + \sqrt 2 \theta \chi^\phi + \theta^2 F^\phi$.}, $\BS$ and $\BPhi$. A stabilizer $\BS$ has a nilpotency of degree two,   $\BS^n = 0$ for $n\geq 2$. This constraint removes the complex scalar $S(x)$, sgoldstino, from the bosonic spectrum.
 The  chiral inflaton multiplet $\BPhi$ has as a  first component  a complex scalar 
$ \Phi(x)= \phi(x) +i b(x)$.
 One can form a  real superfield $\BB\equiv {1\over 2 i}\Big ( \BPhi-\BPhibar\Big)$ with the first component $b(x)$, the sinflaton,  and impose the orthogonality constraint  $\BS \BB=0$. As a result,  fields in the inflaton multiplet are no longer independent: the sinflaton $b(x)$, inflatino $\chi^\phi$ and the auxiliary $F^\phi$ become functions of the spin 1/2 field $\chi^s$ in the $S$-multiplet. All of these fields   vanish in the unitary gauge $\chi^s=0$, which fixes the local supersymmetry of the action.

It follows from $\BS^2 = 0$ and $\BS \BB=0$ that $\BB$ is nilpotent of degree 3, $\BB^m = 0$ for $m\geq 3$. The second real superfield which can be formed from the chiral inflaton superfield is $\BA= {1\over 2} \Big (\BPhi+\BPhibar\Big)$. It starts with one real inflaton scalar field, $\phi(x)$.

The unusual property of these models is that in the unitary gauge, fixing local supersymmetry, there is only one real scalar $\phi(x)$, a massless graviton and a massive gravitino. There is no sgoldstino, no sinflaton and no inflatino. An essential property of these models is that the form of the potential is different from the standard supergravity potentials, as shown in \cite{FKT}:  
\be
V= e^K ( |D_S W|^2 -3 W^2)\Big |_{S=\bar S= \Phi-\Bar \Phi=0} \ .
\ee
 First,  the 3 scalars $S$, $\bar S$ and $\Phi-\Bar \Phi$ vanish,   there is no need to stabilize them. Secondly, 
 the terms which would normally be  present in the potential, quadratic and linear in $D_\Phi W$, are absent, despite the fact that $D_\Phi W$ can be arbitrary. This is because the auxiliary field $F^\phi$ from the inflaton multiplet is fermionic as a consequence of  the orthogonality constraint $\BS \BB=0$.
 
Here we will study supergravity models with constrained superfields
\be
K( \BS\, , \BSbar;  \BPhi ,  \BPhibar) ,    ~~ W= f(\BPhi)  \BS + g (\BPhi) ,  ~~ \BS^2= \BS \BB = 0.
\label{KWgen}
\ee 
The consistency of these models with the constraints was studied in \cite{FKT} where it was shown that the superfield \K\, potential can be also brought to the form
\be
K( \BS\, , \BSbar;  \BPhi ,  \BPhibar)=\BS \,  \BSbar + h(\BA) \BB^2 \ .
\label{KWgen1}
\ee 
This means, in particular that the   \K\, potential in supergravity vanishes when  the bosonic constraints are imposed:
\be
K( S \bar S;  \Phi ,  \bar \Phi)\Big | _{S=\bar S =\Phi-\bar \Phi=0} =0 \ .
\label{K0}\ee
Note, however, that  the bosonic constraints   $S=\bar S =\Phi-\bar \Phi=0$,  when deriving the supergravity action, have to be applied only after the relevant derivatives over $S$ and $\bar S$ and $\Phi$ and $\bar \Phi$ are taken.

For models with orthogonal nilpotent superfields,  the  inflaton action is very simple.  Taking into account \rf{K0} and with a proper normalization of S, we find 
 \begin{align}
 e^{-1}{\cal L}_{\rm infl}  = -K_{\Phi, \bar \Phi}  \partial \Phi \partial \bar \Phi+  [3 {g^2(\Phi) } -  f^2 (\Phi) ]\Big | _{\Phi=\bar \Phi} \ .
\label{actionUGflat}
\end{align}

\noindent {\bf \K\, potentials  in models with $ \BS^2= \BS \BB = 0$.}
Many successful inflationary models in supergravity are based on theories where the \K\ potential either vanishes along the inflaton direction, or can be represented in such form after some \K\ transformations, see for example \cite{Goncharov:1983mw,Kawasaki:2000yn,Kallosh:2010ug,Kallosh:2014xwa,Kallosh:2015zsa,Carrasco:2015uma,Carrasco:2015rva,Carrasco:2015pla}. 
In models with $ \BS^2= \BS \BB = 0$, where $B=(\BPhi- \BPhibar)/(2i)$, this requirement is naturally satisfied \rf{KWgen1},  \rf{K0}.

Here we  study the cosmological models with orthogonal nilpotent superfields \rf{KWgen} over several different \K\, potentials.
The simplest \K\ potential with a flat direction describing a canonically normalized inflaton field $\phi = {\rm Re \,\Phi}$  is given by \cite{Kawasaki:2000yn,Kallosh:2010ug}
\be\label{canon}
K = {1\over 4} (\Phi-\bar\Phi)^{2} + S \bar{S} \ .
\ee
Here the geometry of the moduli space is flat.

We will be especially interested in the \K\ potentials for a broad class of cosmological attractors describing Escher-type hyperbolic geometry \cite{Kallosh:2015zsa,Carrasco:2015uma}
 of the inflaton moduli space. Compatibility of the constraints $ \BS^2= \BS \BB = 0$  with the hyperbolic geometry is demonstrated in the Appendix.
Examples of such \K\ potentials include
\be
K= -{3\over 2}   \alpha \log \left[{(1- \Phi\bar\Phi)^2\over (1-\Phi^2) (1-\overline \Phi^2)}  \right] +S\bar S\,      \ .
\label{KdiskNewDisk}\ee
It describes hyperbolic geometry in disk variables.
The same geometry can be described in half-plane variables by the \K\ potential 
 \be
K= -{3\over 2}\, \alpha  \log \left[ {(\Phi + \bar \Phi )^2 \over  4 \Phi \bar \Phi}   \right] + S\bar S\,     .
\label{KhalfNewHalf}\ee
These two versions correspond to  equivalent ways of describing the \K\ geometry of $\alpha$-attractors. See refs.~\cite{Carrasco:2015uma,Carrasco:2015rva,Carrasco:2015pla} 
for a detailed discussion of this issue.\footnote{The third equivalent choice 
corresponds to the choice of \K\, potential made in \rf{canon} when the nilpotency constraint $\BB^3=0$ is taken into account. It is related to \rf{KdiskNewDisk} and \rf{KhalfNewHalf} by a change of variables  \cite{Carrasco:2015uma,Carrasco:2015rva,Carrasco:2015pla},  see also Appendix.}
 
One may also consider these  \K\ potentials with the term $S\bar S$ under the logarithm. In all of these cases, the \K\ potential  vanishes for $\Phi = \bar\Phi$, and $K_{S, \bar S} = 1$ or can be brought to $K_{S, \bar S} = 1$ by a holomorphic transformation defined in \cite{FKT}. The inflaton action is given by \rf{actionUGflat}, and the inflaton potential is given by a simple expression
\be \label{pot}
V = f^{2}(\phi) - 3g^{2}(\phi) \ .
\ee 
 
This result is similar to the expression $V = f^{2}(\phi)$ for the family of models with $W = S f(\Phi)$ developed in \cite{Kallosh:2010ug}. The new generation of models is different in two respects. First of all, it describes a non-vanishing gravitino mass 
\be
m_{3/2}(\phi) = g(\phi) \ .
\ee
Additionally, it may also describe non-vanishing vacuum energy (cosmological constant) at the minimum of the potential. Without any loss of generality one may assume that the minimum of the potential corresponding to our vacuum state is at $\phi = 0$. The cosmological constant is equal to 
\be
\Lambda = f^{2}(0) - 3g^{2}(0) \ .
\ee
The condition that $\phi = 0$ is a minimum implies that 
$
f'(0) = \sqrt 3 g'(0),
$
up to small corrections vanishing in the limit $\Lambda \to 0$.

These  conditions, plus the requirement that the functions $f(\phi)$ and $g(\phi)$ are holomorphic,  leave lots of freedom to describe observational data. Indeed there are many ways to do so, depending on the choice of the \K\ potential.

Even though the expression of the potential $V = f^{2}(\phi) - 3g^{2}(\phi)$ is valid for all choices of the \K\ potentials described above, the field $\phi$ in the theories with the \K\ potentials \rf{KdiskNewDisk} and \rf{KhalfNewHalf} is not canonically normalized. In the theory  \rf{KdiskNewDisk} the field $\phi $ is related to the canonically normalized inflaton field $\vp$ as follows:
\be
\phi = \tanh{\vp\over \sqrt{6\alpha}} \ .
\ee
Meanwhile for the theory  \rf{KhalfNewHalf} one has
\be\label{star}
\phi = e^{-\sqrt{2\over 3\alpha}\vp} \ .
\ee
Thus, the potential $V = f^{2}(\phi) - 3g^{2}(\phi)$,  expressed in terms of a canonically normalized field $\vp$, depends on the choice of the \K\ potential.  In the next section we will describe several realistic inflationary models in this context.

\vskip 8pt 
\noindent {\bf   Inflationary models.}

\vskip 5pt


\noindent {{\it   \underline{Model 1}:} {$f(\phi) = M\phi^{2} + a ,  ~~~ g(\phi) = b$}.}
\vskip 3pt
\noindent The potential in this model is 
\be\label{canpot}
V = M^{2}\phi^{4} + 2a M\phi^{2} +a^{2}-3b^{2} \ .
\ee
The cosmological constant in this model, and all other models we present here, is equal to 
\be
\Lambda = a^{2}-3b^{2}.
\ee
In realistic models we should have $\Lambda \sim  10^{-120}$ due to an almost precise cancellation between $a^{2}$ and $3b^{2}$ in accordance with a string landscape scenario. The gravitino mass is $m_{3/2} = b$, 
which nearly coincides with $a/\sqrt 3$. For $b \sim 10^{-15}$ one can have the gravitino mass in the often discussed TeV range. To have a proper amplitude of scalar perturbations one should have $M \sim 10^{{-5}} \gg a,\ b$.

If we consider a model with the simplest canonical \K\ potential \rf{canon}, the potential \rf{canpot} is quartic with respect to the canonically normalized inflaton field, which  rules out this simple model.

The situation instantly improves in the theory  with the logarithmic \K\ potential \rf{KdiskNewDisk}, which yields the following  potential in terms of the canonically normalized field $\vp$:
\be\label{noncanpot}
V = M^{2} \tanh^{4}{\vp\over \sqrt{6\alpha}} + 2a M \tanh^{2}{\vp\over \sqrt{6\alpha}} +a^{2}-3b^{2} \ .
\ee
This is the typical T-model $\alpha$ attractor potential \cite{Kallosh:2013hoa}. Inflation occurs at the plateau where $\tanh {\vp\over \sqrt{6\alpha}} \approx 1$. In this regime the second term in \rf{noncanpot} is much smaller than the first term, and both terms are much greater than $\Lambda$, so inflation is  described by the quartic T-model potential 
\be\label{noncanpotT}
V = M^{2} \tanh^{4}{\vp\over \sqrt{6\alpha}} \ .
\ee
The observational predictions of this model for $\alpha \lesssim 10$ practically coincide with the predictions of the simpler model $V = M^{2} \tanh^{2}{\vp\over \sqrt{6\alpha}}$, for the same number of e-foldings $N$  \cite{Kallosh:2013hoa}. However, at the end of inflation in the model  \rf{noncanpotT} the inflaton field begins to oscillate in the approximately quartic potential $\sim \varphi^{4}$. The average equation of state during this stage is the same as of the hot plasma, $p = \rho/3$, as if reheating finishes immediately after inflation. This increases the required number of e-foldings by $\Delta N \sim 3$ \cite{Bezrukov:2011gp}. In its turn, this leads to a slight increase of the spectral index $n_{s}$, which may provide even better fit to the recent Planck data.

\vskip 5pt



\noindent{{\it \underline{Model 2}:} $f(\phi) = M\phi^{2} + a$,~~ $g(\phi) =m\phi^{2} + b$}
\vskip 3pt
\noindent The potential is 
\be
V = (M^{2}-3m^{2})\phi^{4} + 2(Ma-3m\, b)\phi^{2} +a^{2}-3b^{2} \ .
\ee

This model is very similar to the previous one, but there is one potentially interesting difference: The gravitino mass depends on the inflaton, 
$
m_{3/2} = m\phi^{2}+ b.
$



\vskip 8pt

\noindent {{\it \underline{Model 3}:}
$f(\phi) =  \sqrt{{M^{2}\over 2}\phi^{2}+a^{2}}$, ~~~ $g(\phi) =  b$}

\vskip 3pt

\noindent The potential is
\be
V = {M^{2}\over 2}\phi^{2}   +a^{2}-3b^{2} \ .
\ee
The potential for $\phi$ is exactly quadratic, plus a cosmological constant. 

In the theory  with the logarithmic \K\ potential \rf{KdiskNewDisk} this potential becomes a potential for the simplest $\alpha$-attractor model  of the canonically normalized field $\vp$:
\be\label{noncanpot1}
V = {M^{2}\over 2}  \tanh^{2}{\vp\over \sqrt{6\alpha}}  +a^{2}-3b^{2} \ .
\ee
The  gravitino mass is  $m_{3/2} =  b$.

\vskip 8pt

\noindent {{\it \underline{Model 4}:} $f(\phi) =  \sqrt{{M^{2}\over 2}\phi^{2}+a^{2}}, ~~g(\phi) = \sqrt{{m^{2}\over 2}\phi^{2}+b^{2}}$}
  \vskip 3pt 
\noindent In this model one has 
\be
V = {M^{2}-3m^{2}\over 2}\phi^{2} + a^{2}-3b^{2} \ .
\ee
In the theory  with the logarithmic \K\ potential \rf{KdiskNewDisk} the potential of a canonically normalized inflaton field  becomes 
\be\label{noncanpot2}
V = {M^{2}-3m^{2}\over 2}  \tanh^{2}{\vp\over \sqrt{6\alpha}}  +a^{2}-3b^{2} \ .
\ee
This model is very similar to Model 3, but the gravitino mass is $\phi$-dependent,
$
m_{3/2} = \sqrt{{m^{2}\over 2}\phi^{2}+b^{2}}.
$

\vskip 8pt

\noindent{{{\it \underline{Model 5}:} $f(\phi) = \sqrt{F^{2}(\phi)+a^{2}}$, ~~$g(\phi) = \sqrt{G^{2}(\phi) +b^{2}}$}
\vskip 3pt
\noindent In this model 
\be
V =  F^{2}(\phi) -3 G^{2}(\phi) + a^{2}-3b^{2}, \quad  m_{3/2} = \sqrt{G^{2}(\phi) +b^{2}}.
\ee
Because of the freedom of choice of the holomorphic functions $F$ and $G$, one can have a wide variety of potentials fitting all observational data even if the fields $\phi$ is canonically normalized, with the \K\ potential \rf{canon}, see e.g.  \cite{Kallosh:2014xwa}.  Meanwhile in the theories with the \K\ potential \rf{KdiskNewDisk} one finds a family of T-model $\alpha$-attractors with 
\be
V =  F^{2}(\tanh {\vp\over \sqrt{6\alpha}})- 3G^{2}(\tanh {\vp\over \sqrt{6\alpha}})   + a^{2}-3 b^{2}  \ .
\ee
For a wide range of functions $F$ and $G$, these theories have universal cosmological predictions for $\alpha \lesssim 10$ and any given number of e-foldings: $n_{s} = 1-2/N$, $r = 12\alpha/N^{2}$  \cite{Kallosh:2013hoa}. However, by a proper choice of the function $F$ one can modify the required number of e-foldings $N$, which can be useful for  tuning  the predictions for $n_{s}$.

\vskip 8pt

\noindent{{{\it \underline{Model 6}:} $f(\phi) = \sqrt{(1-\phi)^{2}+a^{2}}$, ~~~ $g(\phi) = b$}
\vskip 3pt
\noindent It is a particular version of Model 5 for $F(\phi) = M(1-\phi)$ and $G(\phi) = 0$. This yields
\be
V =  M^{2}(1-\phi)^{2}  + \Lambda, \quad \Lambda =  a^{2}-3 b^{2}, \quad m_{3/2} = b .
\ee
Using the half-plane \K\ potential \rf{KhalfNewHalf} and the relation 
$\phi = e^{-\sqrt{2\over 3\alpha}\vp}$ \rf{star} one finds
\be
V =  M^{2}\left(1-e^{-\sqrt{2\over 3\alpha}\vp}\right)^{2}  + \Lambda   \ .
\ee
This represents the family of E-model $\alpha$-attractors \cite{Kallosh:2013hoa,Kallosh:2015zsa}, which reduces to the  Starobinsky model for $\alpha = 1$, $\Lambda = 0$ and $m_{3/2} = 0$. Meanwhile our class of  theories describes E-models with arbitrary $\alpha$,  $\Lambda$ and  $m_{3/2}$.

\noindent {\bf Conclusions. }
As one could see from the previous section, it is very easy to formulate and analyze  models with orthogonal nilpotent fields. Previously, it was a much more complicated task because of certain constraints  imposed on inflationary models with nilpotent fields, see e.g.   \cite{Kallosh:2014via,Dall'Agata:2014oka,Kallosh:2014hxa} and the more advanced models presented in  \cite{Carrasco:2015pla}. These constraints where required for simplification of the fermionic sector, but they are no longer required in the new class of models where the fermionic sector is trivial because the inflatino disappears in the unitary gauge. As a result, we have lots of flexibility in finding  economical models containing only inflaton, graviton and gravitino, and yet capable of  simultaneously describing inflation, dark energy and SUSY breaking.

The absence of the inflatino also helps us argue that there there is no problem with the unitarity bound during inflation in this class of models. 
The effective cutoff  in supergravity is the scale at which scattering amplitudes violate unitarity bound.  In the theories with nilpotent fields, this cutoff is expected at  $\Lambda \simeq (H^2+ m_{3/2}^2)^{1/4} > \sqrt{H}$, in  units $M_{p}=1$ \cite{Kallosh:2000ve,Dall'Agata:2014oka,Kahn:2015mla,FKT}. During inflation with $H\ll 1$, the UV cut-off $\Lambda > \sqrt{H}$ is much higher than the typical energy of inflationary quantum fluctuations $\sim H$. In general, there could be some additional contributions to scattering due to gravitino-inflatino mixing, but in the theory that we consider there is no inflatino, and therefore  no violation of the unitarity  bound is expected during inflation at sub-Planckian energy density. 

We hope to return to this issue in the future, simultaneously with investigation of reheating in the new class of models. In particular, we expect that the absence of the inflatino should significantly simplify the theory of  non-thermal gravitino production by an oscillating inflaton field \cite{Kallosh:1999jj,Kallosh:2000ve,Giudice:1999yt,Nilles:2001ry}.

\vskip 5pt

We are grateful to S. Ferrara and J. Thaler for many enlightening  discussions and collaboration. JJMC is supported by the European Research Council under ERC-STG-639729, `Strategic Predictions for Quantum Field Theories'.    The work of RK  and AL is supported by the SITP, and by the NSF Grant PHY-1316699.  The work of AL is also supported by the Templeton foundation grant `Inflation, the Multiverse, and Holography.'


\appendix
\section{Appendix A. Hyperbolic geometry models with orthogonal nilpotent superfields}
Three equivalent versions of $\alpha$-attractor models with $S^2(x, \theta)=0$ and  hyperbolic geometry of the inflaton moduli space \cite{Kallosh:2015zsa,Carrasco:2015uma} are given either by a  disk geometry $
 Z\bar Z<1$,
\bea
&&K= -{3\over 2}   \alpha \log \left[{(1- Z\bar Z)^2\over (1-Z^2) (1-\overline Z^2)}  \right] +S\bar S\ , \nonumber\\  
&&\,   \qquad \qquad W= A(Z)  + S B(Z)\, ,  \quad 
\label{KdiskNew}\eea
or  a half-plane geometry, $T+\bar T >0 $,
 \bea
&&K= -{3\over 2}\, \alpha  \log \left[ {(T + \bar T )^2 \over  4 T \bar T}   \right] + S\bar S\, , \nonumber\\ &&  \qquad W= G(T)+ S F(T)\ .
\label{KhalfNew}\eea
or a Killing adapted geometry
\bea
  K&=& -3\alpha \log \Big [\cosh {\Phi-\bar \Phi \over \sqrt{6\alpha}} \Big] + S \bar{S}\, , \cr    W&= &A( \tanh {\Phi\over \sqrt {6\alpha}})  + S B( \tanh {\Phi\over \sqrt {6\alpha}}) \nonumber \\
 &=& G(e^{ \sqrt{2\over 3\alpha}  \Phi })+ S F(e^{ \sqrt{2\over 3\alpha}  \Phi })\ . 
\label{JJ}\eea
In all of these  models the \K\, potential and the superpotential, separately,  are related by a change of variables 
\be
T= {1+Z\over 1-Z}, \quad  Z= \tanh {\Phi\over \sqrt {6\alpha}} ,  \quad T= e^{ \sqrt{2\over 3\alpha}  \Phi }  .
\label{Kil}\ee
Now we would like to impose the orthogonality constraint on our superfields. The Killing-adapted variable $\Phi$ in \rf{JJ} is an unconstrained superfield whose scalar part is not restricted by the boundaries. Therefore we may start by imposing the orthogonality and the nilpotency constraint in the form
\be
\BS^2=0 , \quad \BS(\BPhi- \BPhibar)=0,\quad (\BPhi- \BPhibar)^n=0,  \ n\geq 3 \ ,
\label{orth}\ee
 Using the relation between these variables, one can derive the related constraints for the disk and half-plane variables $Z$, and $T$, respectively. 
 \be 
  \BS (\BPhi- \BPhibar)=0   \quad \Rightarrow \quad  \BS (\BZ- \BZbar)=0   \quad \Rightarrow \quad \BS (\BT- \BTbar)=0 .\nonumber
\ee 
It can be shown also that the  \K\, potentials  in \rf{KdiskNew}, \rf{KhalfNew}, \rf{JJ},  take the form of eq. \rf{KWgen1} due to orthogonality constraint.


\end{document}